# Optical Phonon Limited High Field Transport in Layered Materials

Hareesh Chandrasekar, Kolla L. Ganapathi, Shubhadeep Bhattacharjee, Navakanta Bhat and Digbijoy N. Nath

*Abstract*—An optical phonon limited velocity model has been employed to investigate high-field transport in a selection of layered 2D materials for both, low-power logic switches with scaled supply voltages, and high-power, high-frequency transistors. Drain currents, effective electron velocities and intrinsic cut-off frequencies as a function of carrier density have been predicted thus providing a benchmark for the optical phonon limited high-field performance limits of these materials. The optical phonon limited carrier velocities of a selection of transition metal dichalcogenides and black phosphorus are found to be modest as compared to their n-channel silicon counterparts, questioning the utility of these devices in the source-injection dominated regime. h-BN, at the other end of the spectrum, is shown to be a very promising material for high-frequency high-power devices, subject to experimental realization of high carrier densities, primarily due to its large optical phonon energy. Experimentally extracted saturation velocities from few-layer $MoS_2$ devices show reasonable qualitative and quantitative agreement with predicted values. Temperature dependence of measured $v_{sat}$ is discussed and found to fit a velocity saturation model with a single material dependent fit parameter.

## I. INTRODUCTION

The family of van der Waal's semiconductors spans the entire gamut of band gaps, and hence functionalities, from narrow-band gap materials (black phosphorus, transition metal dichalcogenides) for low-power switching, to wide band gap materials (hexagonal boron nitride) for high-power devices. This therefore offers the possibility of extreme band gap engineering, not viable in other material families. Taken in conjunction with the ease of integration offered by van der Waal's epitaxy,[1] this enables the possibility of fabricating high power amplifiers/switches alongside low-power, high-speed control circuitry on the same chip. Transistors of layered materials are being explored for flexible electronics and as potential alternatives to silicon due to the possibility of ultimate 2-D electrostatic control over the channel thus enabling continued scaling of device dimensions. In highly scaled devices, it is the source-injection velocity more than channel mobility, which determines parameters of transistor performance such as transconductance and cut-off frequency ($f_T$). While low-field transport and mobility of various layered materials such as black phosphorous (BP),[2] transition metal dichalcogenides – $MoS_2$, [3, 4] $WS_2$, $MoSe_2$[4] $WSe_2$[4] and their alloys such as $Mo_xW_{1-x}S_2$[5] – have been studied in some detail, in addition to the transistor performance of these materials in the ballistic regime[6-8], the high-field transport in these materials in the non-ballistic regime, which is of utmost importance under normal operating conditions, remains unexplored. Also, as will be shown in this work, the strength of electron-phonon interactions prevents ballistic transport of carriers, except in very highly scaled channels of these materials. In this paper we investigate the efficacy of narrow and wide band gap layered 2D materials for low-power logic switches with scaled supply voltages and high-power transistors respectively, under optical phonon scattering limits.

## II. DESCRIPTION OF THE OPTICAL PHONON MODEL

Interaction with optical phonons is the dominant scattering mechanism for carrier transport in the non-ballistic regime for many layered semiconductors at room temperatures.[9] In this report, we adopt the optical-phonon limited scattering model proposed by Fang et al.,[10] in order to predict the high-field performance limits of transistors for a selection of such channel materials for electron transport, namely multi-layers of $MoS_2$, $WS_2$, $MoSe_2$, $WSe_2$ and black phosphorus (BP) for logic switching, and hexagonal boron nitride for high-power high-frequency applications. Such a model has been successfully employed previously to explain unusual features in highly scaled GaN high electron mobility transistors.[10] The formalism of this model is described in brief as follows. The interplay between source injection and backscattering of carriers at the source-channel barrier determines the carrier concentrations in the channel and hence critical transistor parameters such as current density, transconductance and cut-off frequency. The energy difference between injected and backscattered electrons, described by their quasi-Fermi levels, is determined by optical phonon emission and is fixed to the optical phonon energy, $E_{op}$ (see Fig. 1). Backscattering only takes place when the energies of injected carriers are higher than the optical phonon energy, i.e. $E_{in}>E_{op}$, as seen from Fig. 1 and for $E_{in}<E_{op}$ all carriers are injected into the channel. The corresponding k-space occupancy hence gives rise to a critical carrier density beyond which backscattering occurs, giving rise to a peak in the current densities and effective electron velocities at this point.

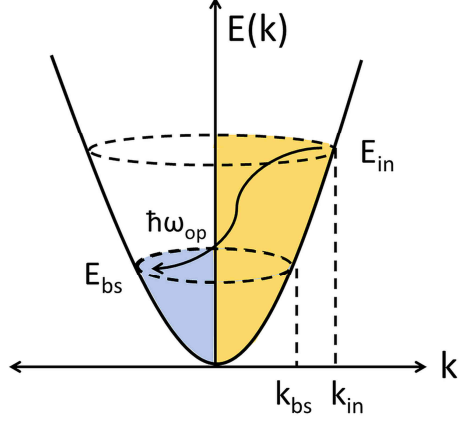

Fig. 1. Energy diagram, at the source injection point, of the optical phonon scattering limited model. The energy and momentum states for injected and backscattered electrons are indicated by $E_{in}$, $k_{in}$ and $E_{bs}$, $k_{bs}$ respectively. Backscattering takes place once $E_{in}>E_{op}$ and the difference in energy between injected and backscattered electrons is given by the optical phonon energy $\hbar\omega_{op}$.

## III. RESULTS AND DISCUSSION

The optical phonon energy in case of GaN is 92 meV and the optical phonon limited mean free path as given by $\lambda_{op} = \alpha^*_B \varepsilon_\infty/(\varepsilon_0 - \varepsilon_\infty)$ is 3.5 nm[10], where $\alpha^*_B$ is the effective Bohr radius, and $\varepsilon_\infty$ and $\varepsilon_0$ are the high-frequency and static dielectric constants. The mean free path for representative layered materials h-BN, $MoS_2$, $WS_2$ and black phosphorus, are listed in Table I, along with their reported dielectric constants and effective masses.

TABLE I

| Material | Dielectric constants $(\varepsilon_0, \varepsilon_\infty)$ | Mean effective mass, $(m_l m_t)^{1/2}$ | Optical Phonon energy, $E_{op}$(meV) | Band Gap $E_g$ (eV) | Mean Free Path, $\lambda_{op}$ (nm) |
|---|---|---|---|---|---|
| h-BN | 7.04, 4.95[11] | 0.26 $m_0$[12] | 169.5[13] | 5.2[13] | 3.4 |
| $MoS_2$ | 7.6, 7[14] | 0.71 $m_0$[15] | 48[16] | 1.29[15] | 6.6 |
| $WS_2$ | 7.0,[17] 5.76[18] | 0.62 $m_0$[15] | 44.6[16] | 1.35[15] | 2.8 |
| $MoSe_2$ | - | 0.64 $m_0$[15] | 35.5[16] | 1.09[15] | - |
| $WSe_2$ | - | 0.56 $m_0$[15] | 31[16] | 1.2[15] | - |
| BP | 12[19], 10[20] | 0.42 $m_0$[2] | 17[20] | 0.36[2] | 7.6 |

The static and high-frequency dielectric constants, mean effective masses, optical phonon energies and optical phonon limited mean free paths for h-BN, $MoS_2$, $WS_2$, $MoSe_2$, $WSe_2$ and BP used in this study.

We see that these values for mean free path, which are comparable to those for GaN, clearly illustrate that optical phonon scattering is expected to play a major role in determining transistor performance of these semiconductors. Since transport in individual layers of these van der Waal's semiconductors can be considered independent of each other, a 2D formalism can be adopted to explain the carrier distribution and velocities. The current per unit width, J, of the transistor after summing over the range of occupied states in momentum space is given by:

$$J=\frac{1}{\sqrt{2\pi}}\frac{qv_{th}kTg\dot{m}^*}{\pi\hbar^2}[\Im_{1/2}(\frac{E_i}{kT})-\Im_{1/2}(\frac{E_i}{kT}-\frac{E_{op}}{kT})] \qquad (1)$$

where $v_{th}$ is the thermal velocity given by $(2kT/m^*)^{1/2}$, g refers to the valley degeneracy of the conduction band, $m^*$ is the effective mass of electrons, $\Im_{1/2}(\frac{E_{in}}{kT})$ is the Fermi-Dirac integral for forward injection given by $2\sqrt{\frac{kT}{\pi}}\int_0^\infty \frac{\sqrt{E}}{1+e^{(E-E_{in})/kT}}dE$, with $E_i$ representing the injection quasi Fermi-level and $\Im_{1/2}(\frac{E_{in}}{kT}-\frac{E_{op}}{kT})$, the Fermi-Dirac distribution of electrons due to backscattering which is lower by the optical phonon energy $E_{op}$. The carrier concentration can be calculated is a similar manner by summing up the injected and backscattered carriers as

$$n_s = \frac{gm^*kT}{2\pi\hbar^2}[\log(1+e^{\frac{E_{in}}{kT}}) + \log(1+e^{\frac{E_{in}-E_{op}}{kT}})] \qquad (2)$$

The drain current densities of a selection of layered materials are evaluated using (1) and (2) for a range of carrier densities from $10^{11}$ cm$^{-2}$ to $5\times10^{13}$ cm$^{-2}$ as shown in Fig. 2 along with a 2DEG in GaN for comparison (see inset).

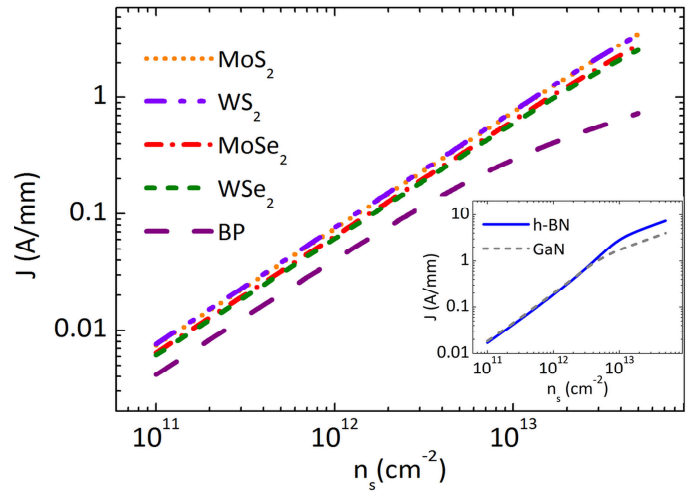

Fig. 2. Log-log plot of normalized drain current (A/mm) in candidate layered materials for low-power switching - MoS$_2$, WS$_2$, MoSe$_2$, WSe$_2$ and black phosphorous (BP) - for a technologically relevant range of carrier densities from $10^{11}$ cm$^{-2}$ to $5\times10^{13}$ cm$^{-2}$. Inset shows the same plot for h-BN, a potential wide bandgap channel material, in comparison with a GaN based 2DEG for high-power applications. Current in MoS$_2$>WS$_2$>MoSe$_2$>WSe$_2$>BP for similar carrier densities, following the trend in their optical phonon energies. Corresponding values for h-BN is higher than other layered materials and is even higher than GaN at higher electron concentrations.

We see that the trend in current carrying capabilities of layered materials follows the trend of their optical phonon energies, monotonically decreasing from h-BN (inset of Fig. 2) to MoS$_2$ down to BP. MoS$_2$ and WS$_2$ exhibit the highest current for the same carrier densities and are hence expected to have the highest on-currents as compared to other low-power candidate materials considered here. h-BN, in addition to having higher current densities than these other layered materials as expected, also outperforms GaN in this regard due to its higher optical phonon energy making it a promising material for high-power transistors. However, this is under the assumption that h-BN is used as a channel material in a device with high carrier densities, which in itself presents significant material and device challenges to be surmounted.

The effective electron velocity at the source, defined by $v_T = \partial J/\partial n_s$ is carrier concentration dependent. The saturation velocity and $v_T$ are similar for silicon MOSFETs while this distinction might be relevant for other materials.[10, 21] However, we see that the low mean free path for electron-phonon interaction in these materials (see Table I) prevents velocity overshoot effects at the source. Therefore, the effective source velocity sets an upper limit for carrier transport and hence the saturation velocity in these devices. Initially, for low carrier densities, optical phonon emission and hence backscattering is blocked due to the lack of available states at energies $E_{in} \leq E_{op}$ (see Fig. 1). At higher carrier densities, a square root dependence is brought about by optical phonon emission which takes place due to the availability of k-states beyond a certain critical electron density($E_{in}>E_{op}$). This gives rise to a maximum in the electron velocity as seen most clearly in the case of h-BN

(Fig. 3(b)). From Fig. 3 (a), it is evident that the source injection velocity for the four transition metal dichalcogenides show a very weak carrier density dependence in this range while the corresponding values for black phosphorous reduce from $2.6\times10^6$ cm/s to $1.26\times10^6$ cm/s for the considered range of carrier densities.

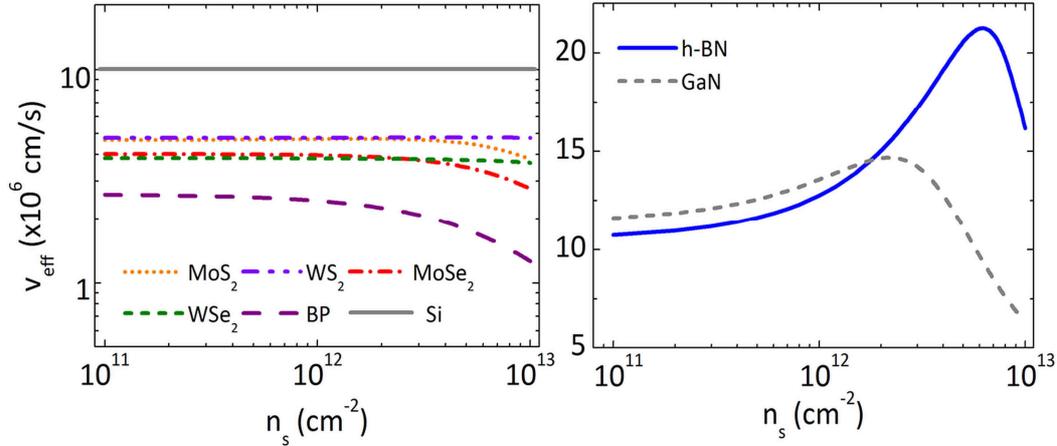

Fig. 3. Effective electron velocity as a function of carrier density for layered materials (a) TMDs and BP for low-power switching, with the saturation velocity of silicon shown for comparison and (b) wide band gap h-BN with the corresponding values of GaN for comparison. The peak in electron velocities occurs due to the transformation to a square-root dependence of current on the carrier density beyond a critical carrier concentration.

As seen from Fig. 3(a), the optical phonon limited electron velocities are highest for $WS_2$ (~ $5\times10^6$ cm/s) followed closely by $MoS_2$ whereas BP has the lowest. We see that these values are at least 2x lesser than the saturation velocity of silicon shown alongside for comparison. h-BN with high charge density, when enabled, on the other hand, would exhibit the highest effective velocities of all layered materials considered here (maximum of $2.1\times10^7$ cm/s at a carrier density of $6\times10^{12}$ cm$^{-2}$) and would also outperform GaN at higher carrier densities illustrating its promise for high-power, high frequency electronics. This is also better illustrated in Fig. 4 where we plot the intrinsic cutoff frequency as given by the transit time of carriers across the channel length of an exemplar 50 nm device, ignoring all other RC contributions to delay, i.e $f_t=v_{sat}/2\pi L_g$.

The intrinsic switching speed of transistors in a circuit depends upon both the source injection velocity, and significantly on the low-field mobility in the channel. The phonon-limited mobility for monolayer $MoS_2$ is predicted to be ~410 cm$^2$/Vs[9] (observed mobilities for bulk multi-layers are ~100 cm$^2$/Vs[4]) as compared to ~300 cm$^2$/Vs electron field-effect mobility for ultra-thin body SOI MOSFETs ($t_{Si}$ = 2.99 nm, electric field = 0.1 MV/cm)[22].

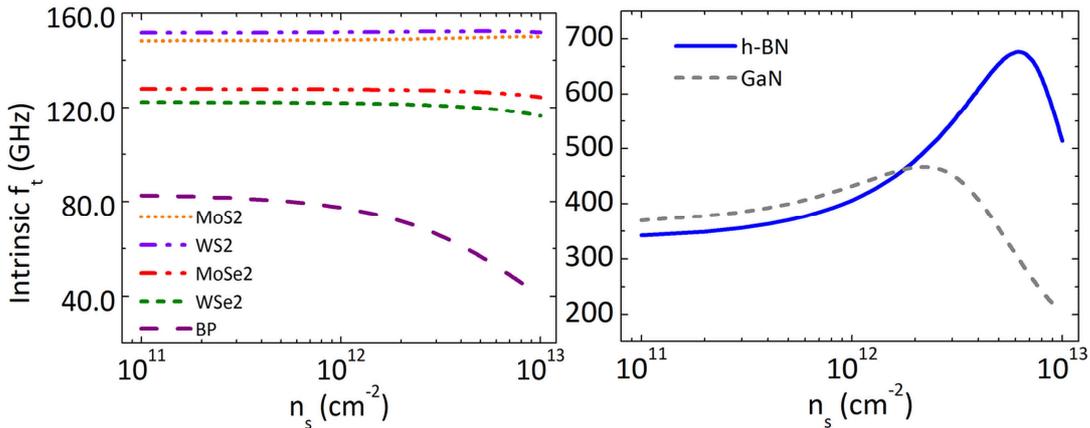

Fig. 4. Intrinsic transistor cut-off frequency $f_t$, as determined by the transit time of carriers across a 50 nm channel, as a function of electron densities for the layered materials (a)TMDs and BP for low-power switching (b)h-BN and GaN for high-power. While h-BN exhibits the highest intrinsic $f_t$ values, we see that TMDs and BP exhibit very modest values of cut-off frequency even for the ideal case all capacitances ignored.

Since the effective electron velocities and mobilities of TMDs are at best comparable to those of silicon, it is clear that the intrinsic switching speeds of few-layer TMDs as n-channel transistors would be lower than their silicon counterparts, even without the challenges involved in reducing contact resistances and appropriate device design to minimize parasitics. While the electron velocity in BP is ~5–8x lower than that of silicon, the higher electron mobility of BP (~1000 cm$^2$/Vs[23]) makes it a viable candidate material for conventional low-power transistors applications. We see that these results offer an interesting guideline for effective device design using layered materials. Transistors of TMDs or BP are better operated at low field regimes, with the attendant advantage of high channel mobilities, as opposed to being biased in the source-injection dominated regime. This consideration hence imposes constraints on the choice of channel lengths and/or supply voltages that can be employed for transistors fabricated from these materials. It is to be noted that the trade-off between the modest effective velocities in these materials and suppression of short-channel effects in highly scaled geometries (such as drain-induced barrier lowering (DIBL), for example)[24] offered by atomic scale confinement of carriers is a factor that needs to be investigated in more detail.

h-BN on the other hand shows promise as a potential candidate in high-power, high-frequency transistors in view of its exceptional current carrying capability and excellent carrier velocities (with a peak intrinsic $f_t$ of 676 GHz, see Fig. 4(b)) with predicted current density and cut-off frequency metrics exceeding those of currently used GaN-based transistors at higher carrier densities. It should be noted that space-charge limited carrier transport in mono-layer boron nitride has only recently been reported[25] and there remain several challenges involved in the practical realization of h-BN transistors ranging from the scalable and facile growth of this material, effective doping and contact engineering schemes to name but a few.

In order to experimentally validate some of the predictions of the optical phonon limited scattering model of carrier transport, we fabricated and characterized transistors of multi-layer MoS$_2$. MoS$_2$ was exfoliated using scotch-tape onto 300 nm SiO$_2$/p++Si substrates followed by acetone treatment to remove the tape residues. 5-6 nm flakes were identified using AFM measurements followed by an appropriate surface-treatment scheme to improve the contact resistances, as reported elsewhere. Electron-beam lithography was employed to pattern source-drain contacts with a separation of 1 μm and Ni was deposited as the contact metal using e-beam evaporation and a lift-off process. The devices were then loaded into a LakeShore variable temperature probe station and all electrical measurements were carried out in a vacuum of at least 1x10$^{-5}$ mbar with temperatures ranging from 100K to 400K. The saturation velocity in these transistors has been extracted using an analytical fit to the family of $I_d$-$V_d$ curves which has previously been applied to Si and MoS$_2$ transistors.[26, 27]

$$I_{DS} = \frac{W}{L}\mu C_{ox}(V_{GS} - V_{th})\frac{V_{DS}}{[1+(\frac{V_{DS}\mu}{Lv_{sat}})^\alpha]^{\frac{1}{\alpha}}} \qquad (3)$$

where $v_{sat}$ stands for the saturation velocity, W and L are the width and length of the transistor, μ is the field effect mobility extracted from the linear region as μ = $g_mL/(WV_{DS}C_{ox})$, $g_m$ the transconductance, $C_{ox}$ the areal oxide capacitance, $V_{th}$ is the threshold voltage as extracted from linear interpolation of the MOSFET transfer curves, $V_{GS}$ and $V_{DS}$ the gate-source and drain-source voltages and α is taken to be 2, which corresponds to the experimentally observed value for electrons.[26]

The saturation velocities thus extracted using (3) for different gate voltages (carrier densities) of the fabricated MoS$_2$ devices are plotted in Fig. 5 alongside theoretical predictions from the optical phonon model as a function of carrier concentrations. We again note that the theoretically predicted source electron velocities set an upper limit on the saturation velocity as the strong electron-phonon interaction checks any velocity overshoot effects that might have been expected otherwise. From Fig. 5 it is evident that the experimental saturation velocities of ~1.6x10$^6$ cm/s are one-third the theoretical values and are comparable to the value of 2.7x10$^6$ cm/s reported in literature.[27] The experimentally extracted values of velocities are expected to be an under-estimation due to the role played by contact resistance in such extraction which has not be considered here.

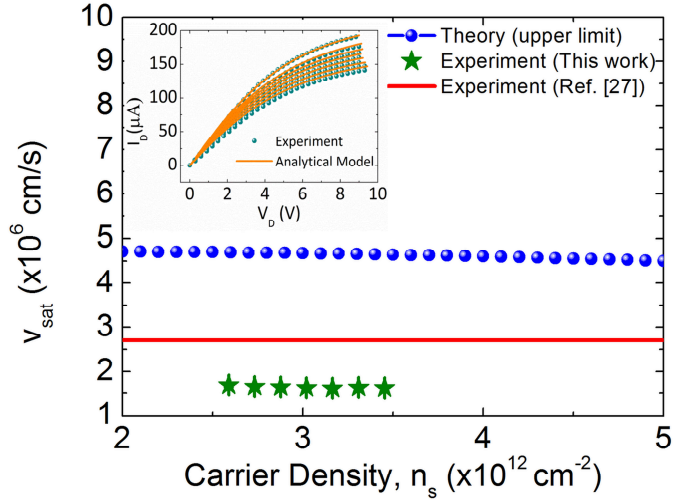

Fig. 5. Saturation velocity of MoS$_2$ transistors as experimentally extracted values (stars) from Eq. (3) for 5-6 nm thick MoS$_2$ flakes of 1 μm gate length and from prior reports (Ref. [27]) for a range of carrier densities. Also shown are the theoretical upper limits (circles) estimated from the optical phonon model. The experimental values are lower by ~3x as compared to the theoretical predictions but displays the same carrier density independent behavior as expected. Inset shows the I$_D$-V$_D$ curves for gate voltages from -12 V to 0 V in steps of 2V and the corresponding fit to the model in (3).

It is noteworthy that the optical phonon model predicts the weak dependence of saturation velocity on the carrier densities for MoS$_2$ and this is exactly what is observed experimentally both in this study as well as prior reports in literature.[27] To further compare the theoretical model to experimental data, I$_D$-V$_D$ measurements were also performed at different temperatures (100K-400K) to estimate the functional dependence of saturation velocity on temperature and the values are plotted in Fig. 6. In the temperature range investigated here, the experimentally extracted saturation velocities v$_{sat}$ shows a good fit to T$^{-0.4\pm0.04}$.

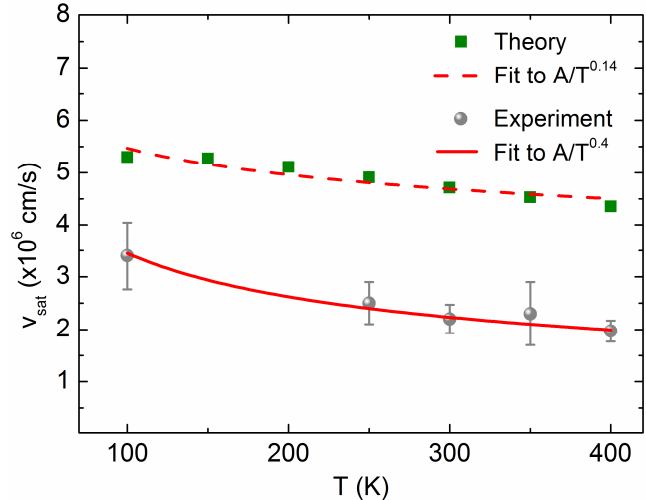

Fig. 6. Temperature dependence of experimental saturation velocities (circles) extracted from MoS$_2$ transistors using (3) as compared to the theoretical upper limits from the optical phonon model (squares). Also shown are fits to a functional form of A/T$^b$ where b=0.14±0.02 and 0.4±0.04 for the theoretical prediction and experimental data respectively.

While the exact temperature dependence of the effective electron velocity as predicted by the optical phonon model is complex, it can be piece-wise approximated in the temperature range considered to a T$^{-0.14\pm0.02}$ fit and is close to the T$^{-0.4\pm0.04}$ observed experimentally. The difference in the exponents for both cases is possibly due to the role of other scattering mechanisms in the fabricated devices as well as the role of contact resistances which has not been de-embedded here. The theoretical T$^{-0.14}$ dependence between 100K to 400K can hence be used to establish an upper limit on experimentally realizable saturation velocities in this temperature range.

## IV. CONCLUSIONS

An optical phonon scattering model has been used to examine some transistor performance metrics, such as currents, saturation velocities and cut-off frequencies, of a selection of layered materials in a 2D formalism. Among the materials considered for low-power switching replacements for silicon, TMDs and BP are found to show poorer electron velocities and cut-off frequencies than comparable silicon n-channel devices. The low-field mobility of electrons in these layered materials, on the other hand, is comparable or greater than in silicon MOSFETs, and hence low-field operation of layered material-based transistors is expected to be more advantageous than biasing them in the source-injection dominated regime. h-BN is shown to be a promising material for high-power, high-frequency electronics exhibiting large current carrying capability, saturation velocity and cut-off frequencies, with a strong carrier concentration dependence predicted for optical phonon limited scattering. Few-layer $MoS_2$ transistors were fabricated to validate the model and experimentally extracted saturation velocities were found to be within the optical phonon limited values and follow the same trends with carrier densities as theoretical predictions. The temperature dependence of the saturation velocity was also analyzed and found to match the functional form of the optical phonon model reasonably well. The results presented here are expected to provide a benchmark for the high-field transistor performance of layered materials.

ACKNOWLEDGMENT

The authors would like to acknowledge the National Nano Fabrication Centre (NNFC) and the Micro and Nano Characterization Facility (MNCF) at the Centre for Nano Science and Engineering for access to fabrication and characterization facilities. H. Chandrasekar thanks Dr. Kausik Majumdar for fruitful discussions.